\documentclass[twoside]{bourlarge}
\usepackage{fayet_h}
\usepackage{color}
\usepackage{amssymb}
\usepackage{amsmath}
\usepackage{graphicx}
\usepackage{cite}
\usepackage{pdfsync}
\usepackage[colorlinks=true]{hyperref}
\hypersetup{pagebackref,urlcolor=blue,citecolor=blue,linkcolor=blue}
\usepackage{url}

\def\be{\begin{equation}}
\def\ee{\end{equation}}
\def\bea{\begin{eqnarray}}
\def\eea{\end{eqnarray}}
\def\ba{\begin{array}} 
\def\ea{\end{array}}
\def\bc{\begin{center}}
\def\ec{\end{center}}

\def\ghost#1{}
\def\sl{\!\!\!/}

\def\simge{\mathrel{%
   \rlap{\raise 0.511ex \hbox{$>$}}{\lower 0.511ex \hbox{$\sim$}}}}
\def\simle{\mathrel{
   \rlap{\raise 0.511ex \hbox{$<$}}{\lower 0.511ex \hbox{$\sim$}}}}

\def\dis{\displaystyle}

\begin{document}
\Large

\title{Scalar Bosons and Supersymmetry}

\author{Pierre {\sc FAYET} \\
 Laboratoire de Physique Th\'eorique de l'\'Ecole Normale Sup\'erieure,\\ 24 rue Lhomond,\\ 75231 Paris cedex 05, France\footnote{and D\'epartement de physique, \'Ecole polytechnique, 91128 Palaiseau cedex, France }}

\maketitle

\begin{abstract}
The recent discovery of a spin-0 Brout-Englert-Higgs boson  leads to further enquire  about other fundamental scalars.
Supersymmetric theories involve, in relation with the electroweak breaking, five such scalars at least, two charged and three neutral ones, 
usually denoted as $H^\pm,\,H,\, h$ and $A$.
They also introduce spin-0 squarks and sleptons as the superpartners of quarks and leptons.

\vspace{1mm}

Supersymmetric extensions of the standard model  lead to the possibility of gauge/BEH 
unification by providing spin-0 bosons as extra states for spin-1 gauge bosons within massive gauge multiplets. 
Depending on its properties the 125 GeV boson observed at CERN
may then also be interpreted, up to a mixing angle induced by supersymmetry breaking,
as {\it the spin-0 partner of the $Z$ under two supersymmetry transformations}, 
i.e. as a  $Z$ that would be deprived of its spin.
\end{abstract}


\section{The electroweak symmetry breaking}

Special relativity and quantum mechanics, operating within quantum field theory, led to the Standard Model of particles and  interactions.
It has met a long series of successes with the discoveries of weak neutral currents (1973), charmed particles (1974-76), gluons mediators of  strong interactions (1979),
 $W^\pm\!$ and $Z$'s mediators of weak interactions (1983), and the sixth quark known as the top quark (1995).

\vspace{2mm}

 Weak, electromagnetic and strong interactions are all understood from the exchanges of spin-1 mediators, $W^\pm$'s and $Z$'s, photons and gluons,  
 between \hbox{spin-$\frac{1}{2}$} quarks and leptons, generically referred to  as the constituents of matter. The $u$ and $d$  quarks are the building blocks for the protons $uud$ and neutrons $ddu$,  and the leptons include the electrons,  muons and taus with their three neutrinos. 
The known fundamental particles are shown in Table \ref{tab:sm0}, without any fundamental spin-0  boson yet.

\begin{table}
\caption{\it The gauge bosons mediators of strong, weak and electromagnetic interactions, and the quarks and leptons, constituents of matter.}
\label{tab:sm0}
\bc
\begin{tabular}{|l|}
\hline  \vspace{-.5mm}\\
\ \ \ spin-1 bosons  {\em \,mediators of interactions}\,:\ \ \ \ \ \ 
{ gluons, \ $W^+\!,\ W^-\!,\,Z$, \,photon,}\ \ \ \ 
\\[3mm] 
\ \ \ spin-$\frac{1}{2}$ fermions  {\em \,constituents of matter}\,: \ \ \ 
$\left\{\ba{c}
  \hbox{6 quarks:}\   \left( \! \ba{c} u\vspace{0mm}\\ d\ea\!  \right)$    $\left(\!  \ba{c} c\vspace{0mm}\\ s\ea\!  \right)$  $\left( \! \ba{c} t \vspace{0mm}\\ b \ea \! \right),
\vspace{3mm}\\
 \hbox{6 leptons:} \  \left(\!\!  \ba{c}\nu_e\vspace{0mm}\\ e^-\ea \!\!\!\right)$     $\left( \!  \! \ba{c} \nu_\mu\!\vspace{0mm}\\ \mu^-\!\ea\!\! \right)$   
  $\left(\!\!  \ba{c} \nu_\tau\!\vspace{0mm}\\ \tau^-\!\ea\! \!\! \right).
 \vspace{1mm}\\
\ea \right.$
\\[-2mm] \\
\hline
\end{tabular}
\vspace{-2mm}{}
\ec
\end{table}

\vspace{2mm}

The spin-1 bosons mediators of interactions are  associated with local gauge symmetries.
The eight gluons mediate the strong interactions, invariant under the color $SU(3)$ gauge group.
The $W^\pm,Z$ and photon, mediators of the electroweak interactions, are associated with 
the $SU(2)\times U(1)$ electroweak gauge group  \cite{ew1,ew2,ws}. It gives very much  the same role to
the  left-handed  quark fields $u_L$ and $d_L$, and similarly  to the left-handed lepton fields  $\nu_{eL}$ and  $e_L$.

\vspace{2mm}
The electroweak symmetry  requires in principle the corresponding spin-1 gauge bosons to be massless.
The weak interactions, mediated by virtual $W^\pm$ and $Z$ production or exchanges,  
would then be long-ranged.
They have instead a very short range $\,\simeq 2\ 10^{-16}$ cm, corresponding to the large masses $m_W\simeq 80$ GeV$/c^2$
and $m_Z\simeq 91$ GeV$/c^2$ of their mediators, almost 100 times the mass of a proton.
The electroweak symmetry should  also require the charged leptons and quarks to be massless, which is not the case.

\vspace{2mm}
Both problems are solved within the standard model through the spontaneous breaking of the electroweak gauge symmetry 
induced by a doublet of complex spin-0 fields $\varphi$  \cite{ws}.
Three of its four real components, instead of being associated with  three unwanted massless  Goldstone bosons \cite{g} as it would be the case if the electroweak  $SU(2)\times U(1)$ symmetry 
were only global, are eliminated by the Brout-Englert-Higgs mechanism \cite{be,h,ghk} 
to provide the additional degrees of freedom required for the $W^\pm$ and $Z$ to acquire masses. 

\vspace{2mm}

The fourth component of the spin-0 doublet, taken as $\phi= \sqrt{2\,\varphi^\dagger\varphi}$, \,adjusts uniformly  in space-time
so that  the potential 
\be
\label{V}
V(\varphi)\,=\,\lambda\ (\varphi^\dagger\varphi)^2-\,\mu^2\,\varphi^\dagger\varphi\,,
\ee
with its famous mexican-hat shape, is minimum, for $\,\phi=v=\sqrt{\mu^2/\lambda}\ $ \cite{ws,g,be,h}. 
\,The electroweak symmetry is then said to be ``spontaneously broken'' (even if $\phi$ itself remains gauge-invariant), meaning by this expression that the $W^\pm\!$ and $Z$ are no longer massless. The local gauge symmetry, which strictly speaking still remains unbroken,  gets now hidden.
The $W^\pm\!$ and $Z$ acquire masses fixed in terms of the electroweak gauge couplings $g$ and $g'$ by
\be
m_W= \hbox{\large$\dis\frac{gv}{2}$}\,,\ \ m_Z=\hbox{\large$\dis \frac{\sqrt{g^2+g'^2}\ v}{2}$}\,=\,{\large\hbox{$\dis \frac{m_W}{\cos\theta}$}}\ .
\ee
The electroweak mixing angle $\theta$ which enters in the definitions of the $Z$ and photon fields
is  fixed by $\tan\theta=g'/g$, the photon staying massless.

\vspace{2mm}
This mechanism of spontaneous symmetry breaking is at the origin of the differentiation between weak interactions, becoming short-ranged, and electromagnetic ones, which remain long-ranged.
The elementary charge $e$ and the Fermi coupling of weak interactions $G_F$ 
are given by $e=g\,\sin\theta$ and  $\,G_F/\sqrt 2=g^2/(8m_W^2)=1/(2v^2)$, so that  
$\,v= (G_F\sqrt 2)^{-1/2}\simeq \,246$ GeV.

\begin{table}[ht!]
\caption{\it Particle content of the standard model. Strong, weak and electromagnetic interactions of quarks and leptons are invariant under the 
$SU(3)\times SU(2)\times U(1)$ gauge symmetry group. The spin-0 BEH boson is associated with the spontaneous breaking of the
electroweak symmetry and the generation of masses for the $W^\pm$ and $Z$, quarks and charged leptons.}
\label{tab:sm}
\bc
\begin{tabular}{|cc|}
\hline & \\[-2mm]
\ \ \ \ spin-1 gauge bosons:\ \ \  &\ \ {\color{black} gluons,}  \  
$ \color{black}W^+\!,\ W^-\!, \, Z, \ \,\hbox{photon}$\ \ \ 
\\[1.5mm] 
spin-{$ \frac{1}{2}$}\ fermions: & {\color{black} 6 quarks} \,+\,
{\color{black}6  leptons} \\[1.5mm] 
 1 spin-0 &
{\color{black} scalar BEH boson}
\\[1.5mm] 
\hline
\end{tabular}
\ec
\vspace*{-4mm}{}
\end{table}

\vspace{2mm}
The introduction of the spin-0 doublet field $\varphi$, called a Brout-Englert-Higgs field or often simply Higgs field, also allows for the charged leptons and quarks to acquire masses, that would otherwise be forbidden by the $SU(2)\times U(1)$ symmetry. \,
Indeed an electron in an electromagnetic field acquires an energy $E = qV$ where $q=-\,e$ is its electric charge and $V$ the electrostatic potential.
The electron field, which interacts with the spin-1 electromagnetic gauge field $A^\mu$, also interacts with the \hbox{spin-0} doublet field $\varphi$, with a coupling constant $\lambda_e$.
The electron is then sensitive to its modulus, i.e. to the physical BEH field (still gauge invariant),
\be
\phi\,=\, \sqrt{2\,\varphi^\dagger\varphi}\ .
\ee

\noindent
It acquires a field-dependent mass parameter $\lambda_e \phi$  and thus, with the BEH field having a constant value 
$\phi=v$ uniform in space-time, a mass $m_e=\lambda_e v$.
\,The same phenomenon occurs  for the other charged leptons $\mu$ and $\tau$, and the quarks. 
They all acquire masses proportional to their couplings to this scalar field $\phi$,
\be
\label{mlq}
m_l = \lambda_l\,v\,,\ \ m_q=\lambda_q\,v\,.
\ee

\noindent
The three neutrinos, which do not interact directly with $\phi$
\,in connection with the absence of right-handed neutrino fields $\nu_R$, remain massless at this stage.
They have, however, very small masses whose origin is as yet unknown, and can oscillate from one flavor to another.

\section{The scalar boson of the standard model}


An immediate consequence of the introduction of a spin-0 field $\varphi$ is that there should be spin-0 excitations, i.e. particles, associated with its quantization. 
The complex electroweak doublet $\varphi$  introduced \cite{ws} to generate a spontaneous breaking of the electroweak symmetry 
\cite{ew1,ew2} describes four field degrees of freedom. Three of them are eliminated by the BEH mechanism \cite{be,h,ghk}, with one only (indeed the one which attracted the least attention at the time)
surviving in the physical theory.

\vspace{1.5mm}

The waves corresponding to the space-time variations of this field $\phi= \sqrt{2\,\varphi^\dagger\varphi}$, once quantized,  are associated with neutral scalar Brout-Englert-Higgs bosons, 
also more commonly referred to as Higgs bosons for historical reasons.
This is very much the same as for electromagnetic waves, whose quanta are the massless spin-1 photons.
The mass of this spin-0 boson, obtained by expanding the potential (\ref{V}) near its minimum, is given by 
\be
\label{mh}
m_h= \,\sqrt{2\mu^2}\,=\,\sqrt{2\lambda v^2}\,.
\ee

This scalar is essential for the consistency of the standard model as a quantum field theory. 
Its couplings to quarks and leptons, obtained from (\ref{mlq}), are proportional to their masses,
\be
\lambda_{q,l}= \,2^{1/4}\,G_F^{1/2}\,m_{q,l}\,.
\ee
Its mass, however,  is not predicted by the theory, but fixed by the quartic coupling $\lambda$ in the scalar potential $V(\varphi)$ in (\ref{V}).

\vspace{2mm}

The possible origin of this $\lambda \,(\varphi^\dagger\varphi)^2$ quartic coupling  is a subject  to which we shall return later, within the framework of supersymmetric theories.
They  lead to consider several spin-0 BEH bosons, charged and neutral, relating their quartic couplings to the squares of the electroweak gauge couplings, $g^2$ and $g'^2$, in particular
through
\be
\lambda = \,\frac{g^2+g'^2}{8}\ .
\ee
They thus also provide,  in particular, a neutral spin-0 BEH boson that would have the same mass as the $Z$ \cite{R},

\vspace{-1mm}
\be
\label{mhZ}
m_h\,=\,\sqrt{2\lambda v^2}\,=\,\hbox{\large$\dis \frac{\sqrt{g^2+g'^2}\ v}{2}$}\,=\,\,m_Z\,\simeq \ 91\ \hbox{GeV}/c^2\,,
\ee
in the absence of supersymmetry breaking effects.

\vspace{2mm}
The scalar boson of the standard model has  long remained its last missing particle 
 after the discovery of the top quark in 1995, escaping until recently all experimental efforts deployed  to detect 
its production, most notably at the  $e^+e^-$ LEP collider at CERN, which established a lower bound of 114 GeV$/c^2$ on its mass \cite{LEP}.

\vspace{2mm}

The existence of a new boson has been established recently, in 2012, at the Large Hadron Collider LHC at CERN \cite{higgs,higgs2}.
This particle, neutral, has a mass close to 125 GeV$/c^2$, and  almost certainly spin 0, rather than 2. It cannot have spin 1 as it is observed to have $\gamma\gamma$ decay modes.
 It shows at this point the properties expected 
from a scalar boson associated with the differentiation between electromagnetic and weak interactions, 
and the generation of masses. 
If it is indeed the scalar boson of the standard model 
this one may now be considered as complete,
this spin-0 particle being its last missing piece (cf. Table \ref{tab:sm}). The standard model would then become the standard theory of particle interactions.

\section{A scalar boson, elementary or not?}

The existence of such a scalar boson has in fact  long been questioned, 
many physicists having serious doubts about the very existence of fundamental spin-0 fields.
Indeed in a theory including very high scales much larger than the electroweak scale, such as a possible grand-unification scale (now believed to be 
of the order of $10^{16}$ GeV), or the Planck scale $\simeq 10^{19}$ GeV possibly associated with quantum gravity,  such \hbox{spin-0} fields tend to acquire very large mass terms, disappearing from the low-energy theory.

 \vspace{2mm}

Many efforts were thus devoted  to replace fundamental spin-0 fields, without much success, by composite fields built from spin-$\frac{1}{2}$ ones, 
e.g. techniquark fields specially introduced for that purpose \cite{tc1,tc2,tc3,tc4}, in view of ultimately avoiding fundamental spin-0 bosons associated with the electroweak breaking.

\vspace{2mm}
One may ask whether the new boson recently found at CERN \cite{higgs,higgs2} is indeed the scalar one of the standard model, or if its properties may deviate from it at some point. 
It is of course not the first spin-0 particle found. Pions, kaons, ...  are also spin-0 particles, but composite  $q\bar q$  states
constructed from quarks and antiquarks. In contrast  the new 125 GeV boson presents at this stage all the characteristics of an elementary particle, the first of its kind. 
Is it alone, or just the first member of a new class?

\section{Introducing supersymmetry}

In the meantime however, the situation concerning our view of spin-0 fields has changed with the introduction of supersymmetry.
Its algebraic  structure \cite{gl,va,wz,revue,martin,ramond} provides a natural framework for fundamental spin-0 fields, now treated on the same footing as spin-$\frac{1}{2}$ ones, 
 to which they are related by supersymmetry transformations.
Indeed, according to common knowledge, supersymmetry is expected to relate
bosons (of integer spin) with fermions (of half-integer spin), as follows:
 \be
 \label{bf}
 \ba{c}
\,\hbox{\color{black} bosons} \ \ \ \ \ 
\stackrel{\stackrel{\hbox{\normalsize supersymmetry}}{\phantom{a}}}
{\hbox{\Large $\longleftrightarrow$}}
\ \ \ \ \ \hbox{\color{black} fermions}\,.
\vspace{0mm}\\
\ \hbox{\small (integer spin)       \hspace{32mm} (half-integer spin)}
\ea
\ee


But can this be of any help in understanding  the real world  of particles and interactions?
If supersymmetry is to act at the fundamental level
the natural idea would be to try to use it to relate the known bosons and fermions in Table \ref{tab:sm0}, or \ref{tab:sm}.
More precisely, can one relate the spin-1 bosons (gluons, $W^\pm\!, \,Z$ and photon) messengers of interactions to the spin-$\frac{1}{2}$ fermions (quarks and leptons) constituents of matter?
This would lead  to a sort of unification
  \be
   \label{bf2}
\ \hbox{ Forces} \ \ \ \ \ 
\stackrel{\stackrel{\hbox{\normalsize supersymmetry?}\!\!\!}{\phantom{a}}}
{\hbox{\Large $\longleftrightarrow$}}
\ \ \ \ \ \hbox{\color{black} Matter}\,. 
\ee
 The idea looks attractive, even so attractive that supersymmetry is frequently presented as a symmetry uniting forces with matter.
 This is however misleading at least at the present stage, and things do not work out that way.
  \vspace{2mm}
 
 Indeed the algebraic structure of supersymmetry  did not seem directly applicable to particle physics \cite{ramond}, in particular as known fundamental bosons
 and fermions do not seem to have much in common.
 There are also other more technical reasons, dealing with the  difficulties of spontaneous supersymmetry breaking, the fate of the resulting Goldstone fermion \cite{fi,for} (even if it is subsequently eaten away by the spin-$\frac{3}{2}$ gravitino \cite{grav}), the 
 presence of self-conjugate Majorana fermions, the requirements of baryon and lepton number conservation, etc..

 \section{Relating bosons and fermions, yes, but how?}

 One has to find out which bosons and fermions might be related under supersymmetry, first considering possible associations between baryons and mesons. Or, at the fundamental level, exploring 
 tentative associations like
\be
 \label{bf3}
\left\{\ \ \ 
\ba{ccc}
\vspace{-6mm}\\ 
 \hbox{photon}\ \ &\stackrel{\hbox{\normalsize ?}}{\longleftrightarrow}&\ \ \hbox{neutrino} 
\vspace{-.8mm}\\ 
 W^\pm\ \ &\stackrel{\hbox{\normalsize  ?}}{\longleftrightarrow}&\ \  e^\pm
\vspace{-.8mm}\\ 
\hbox{gluons}\ \ &\stackrel{\hbox{ \normalsize ?}}{\longleftrightarrow}&\ \ \hbox{quarks} \vspace{-.5mm}\\ 
& ...& 
 \vspace{-.8mm}\\ 
\ea \right.
\ee
But we have no chance to realize in this way systematic associations of known fundamental bosons and fermions.
This is also made obvious from the fact that we know 90 fermionic field degrees of freedom for the quarks and leptons
(for 3 families of 15 chiral quark and lepton fields)
as compared to 28 only  for bosonic ones (16 + 11 + 1 for the gluons, electroweak gauge bosons and the new scalar). In addition these fields have different gauge and 
$B$ and $L$ quantum numbers, preventing them from being directly related.

 \vspace{2mm}
 In supersymmetry we also have to deal with the systematic appearance of self-conjugate Majorana fermions, while Nature seems to know Dirac fermions
 only (with a possible special exception for neutrinos having Majorana mass terms).
 How can we obtain Dirac fermions, and attribute them conserved quantum numbers like $B$ and $L$?

 \section{The need for superpartners}

  To face all these difficulties, we were led to introduce a color octet of spin-$\frac{1}{2}$ \linebreak Majorana fermions \cite{ssm}
 called {\it gluinos} \cite{grav}, although their consideration was at the time forbidden by the general principle of triality \cite{tria}. 
 This one, however,  gets systematically violated within supersymmetric theories, in which it no longer applies. 
 A strict application of this principle would have prevented us from discussing supersymmetric theories.
 
 \vspace{2mm}
 
 As gluons were associated with gluinos, the photon had to be associated, not with any the known neutrinos $\nu_e$ and $\nu_\mu$, or later $\nu_\tau$, 
 but with a ``photonic neutrino'' called the {\it photino}. It is the first member of a larger family of {\it neutralinos}, 
 obtained from mixings of neutral spin-$\frac{1}{2}$ gaugino and higgsino fields, respectively associated with
 gauge and BEH fields under supersymmetry.

 \vspace{2mm}
 But, as far as we know at the moment, baryon and lepton numbers $B$ and $L$ are carried by fundamental fermions only, quarks and leptons, not by bosons.
 They are thus even referred to as ``fermionic numbers''. In a supersymmetric theory however it gets impossible to have $B$ and $L$ carried just by fermions, 
 and not by bosons. But attributing ``fermion number'' to bosons looks like a non-sense\,! 
 To include the standard model within a supersymmetric theory we also have to accept the unconventional idea that a significant number of fundamental bosons 
 may have to carry  baryon or lepton numbers.

 \vspace{2mm}
 These new bosons carrying $B$ or $L$ are now well-known as {\it squarks} and {\it sleptons}.  Their denomination makes apparent and even ``obvious''  
 that they have to carry the same $B$ and $L$  as their fermionic counterparts.

 \vspace{2mm}
 Still this does not guarantee yet that $B$ and $L$ will systematically remain conserved as  observed in Nature,
at least to a sufficiently good approximation. In particular we do not want the proton to undergo too fast decays, as its lifetime should be larger than
about $10^{32}$ years or so.
This requires the consideration of an additional symmetry, namely $R$ symmetry or its discrete version, $R$-parity
\cite{R,ssm,ff}.

 \vspace{2mm}
This one, originally obtained as the parity of a continuous quantum number $R$ carried by the supersymmetry generator, $R_p=(-1)^R$,
is simply $+1$ for all standard model particles  and $-1$ for their superpartners, 
\be
R_p = \left\{\ba{cc}  +1& \hbox{for \it quarks and leptons, gauge and BEH bosons,}
\vspace{2mm}\\
 -1&\hbox{for \it squarks and sleptons, gluinos, charginos and neutralinos.}
 \ea\right.
\ee
It may be rewritten in terms of the spin, baryon and lepton numbers of particles, as 
\be
R_p\,=\, (-1)^{2S}\ (-1)^{3B+L}\,.
\ee
This illustrates its connection with $B$ and $L$, or simply $B-L$, conservation laws, even allowing for $\Delta L= \pm 2$ processes  as in the presence of neutrino Majorana mass terms.

 \vspace{2mm}
 Although one may consider that $R$-parity conservation is not necessary and may thus be questioned \cite{rpv},
its requirement is natural, and its absence usually the source of various troubles. Indeed $R$-parity  prevents direct exchanges of squarks or sleptons between ordinary quarks and leptons, 
that could induce proton decay at a much too high rate.

 \vspace{2mm}
$R$-parity also plays a crucial role in the stability of the lightest supersymmetric particle, or LSP, in general considered to be a {\it neutralino}.
The pair-production (and subsequent decays) of supersymmetric particles should ultimately lead to two unobserved neutralinos, the famous Òmissing energy-momentumÓ signature often used to search for supersymmetry.
Stable massive neutralinos having survived annihilations also turn out to be natural candidates for the non-baryonic dark matter of the Universe.

\begin{table}[tb]
\vspace{-2.5mm}
\caption{\it Minimal content of the Supersymmetric Standard Model (MSSM).
Neutral gauginos and higgsinos mix into a photino, two zinos and a higgsino, further mixed into four neutralinos.
Ordinary particles from the standard model, including additional BEH bosons, in blue, have $R$-parity $+1$.
Their superpartners, in red, have $R$-parity $-1$.
\label{tab:mssm}}
\vspace{1mm}
\begin{center}
\begin{tabular}{|c|c|c|} 
 \hline  &&\\ [-2.5mm] 
 Spin 1       &Spin 1/2     &Spin 0 \\ [1.2mm]\hline  
\hline 
&&\\ [-1.8mm]
\color{blue}gluons       	 &\color{red}gluinos ~$\tilde{g}$        &\\[.8mm]
\color{blue}photon           &\color{red}photino ~$\tilde{\gamma}$   &\\ [.3mm]
--------------&$- - - - - - - - -$&-------------------------------- \\ [-2.2mm]
 

$\begin{array}{c}
\\ \color{blue} W^\pm\\ [1.1mm]\color{blue}Z \ \  \\ [.8mm]
\\ \\
\end{array}$

&$\begin{array}{c}
\color{red}\hbox {winos } \ \widetilde W_{1,2}^{\,\pm} \\ [.8mm]
\color{red}\,\hbox {zinos } \ \ \widetilde Z_{1,2} \\ [1.3mm]
\color{red}\hbox {higgsino } \ \tilde h
\end{array}$

&$\hspace{-12mm}\left. \begin{array}{c}
\color{blue}H^\pm\\ [1.3mm]
\color{blue} h \\
[1.3mm]
\color{blue}H, \,A
\end{array}\ \right\} 
\begin{array}{c}\!\! \hbox {BEH bosons}\\ 
\end{array}\hspace{-12mm}$  \\ &&\\ 
[-4.8mm]
\hline &&

\\ [-2.8mm]
&\color{blue}leptons ~$l$       &\color{red}sleptons  ~$\tilde l$ \\[.8mm]
&\color{blue}quarks ~$q$       &\color{red}squarks   ~$\tilde q$\\ [-2.2mm]&&
\\ \hline
\end{tabular}
\ec
\vspace{+0.5mm}
\end{table}

   \vspace{2mm}

Supersymmetry thus does not relate directly known bosons and fermions. 
All known particles should be associated with new superpartners which appear as their {\it images under supersymmetry}, according to
\be
 \label{bf4}
\left\{\ 
\begin{tabular}{ccc}
{ known bosons } & $\longleftrightarrow$& {\ \ new fermions},\ \ \\[1mm]
{\ \ known fermions} \ \ & $\longleftrightarrow$& { new bosons}.
\end{tabular} \right.
\ee
This was long mocked as a sign of the irrelevance of supersymmetry.
But times have changed,
to the point that supersymmetry gets now frequently referred to as a symmetry which postulates the existence, for each particle of the standard model, of a supersymmetric 
partner differing by 1/2 unit of spin.


 

 \section{The supersymmetric standard model, with its extra spin-0 bosons}

 The resulting supersymmetric standard model involves
 spin-0 squarks and sleptons, and spin-$\frac{1}{2}$ gluinos, charginos (called winos in Table {\ref{tab:mssm}) and neutralinos  \cite{ssm,R,ff,fayet79}. 
 As of today, howewer, they remain unseen after more than three decades of experimental searches, 
 starting in the late seventies. 
 The search for these new particles is now one of the main objectives of the LHC collider at CERN \cite{susyat,susycms}, 
whose energy should soon increase from 8 to 13 TeV. 

\vspace{2mm}

Will this be sufficient? At which energy scale should the new supersymmetric
particles be found?
Is it of the order of the TeV scale, not too far from the electroweak scale and accessible at LHC? Or possibly significantly larger, as it could happen in theories with extra space dimensions
 \cite{gutbis,epjc}, with the mass scale of the new superpartners 
fixed by the (or a) compactification scale $\propto \hbar/Lc$?  \,($L < 10^{-17}$ cm corresponding to $ \hbar c/L > 2$ TeV.)

\vspace{2mm}
 
 In any case the supersymmetric standard model  requires for the electroweak breaking, not a single doublet $\varphi$ as in the standard model, 
 but two at least, 
\be
h_1=\,\left(\!\ba{c}h_1^0 \vspace{.5mm} \\ h_1^- \ea\!\right)  \ \ 
\hbox{and}\ \ \  h_2=\, \left(\!\ba{c}h_2^+ \vspace{.5mm} \\ h_2^0 \ea\!\right)  .
\ee

\vspace{-1mm}

\noindent
They are needed  to construct
\vspace{-.5mm}
 two massive Dirac charginos (also called winos in\break Table \ref{tab:mssm})
from charged gaugino 
($\widetilde W_{L+R}^{\,-}$\,) and higgsino 
 \,($\,\tilde h_{1\,L}^{\,-}$ and $\,(\tilde h_{2\,L}^{\,+})^c$\,) components, without getting stuck with a massless chiral charged fermion \cite{R,ssm}.
These two doublets
allow for the generation of charged-lepton and down-quark masses from $h_1$, 
and up-quark masses from $h_2$. This applies to the various versions of the supersymmetric standard model, from the minimal one known as the MSSM, to others that may include
an extra singlet coupled to the two doublets $h_1$ and $h_2$ as in the N/nMSSM, or USSM if an extra $U(1)$ symmetry is gauged, GMSB models, etc..
 
 \vspace{2mm}

$h_1$ and $h_2$ represent altogether eight real spin-0 fields, among which three get eliminated when the $W^\pm$ and $Z$ acquire masses.  
This  results in  five spin-0 BEH scalars, two charged and three neutrals, usually denoted as $H^\pm,\,H,\, h$ and $A$, also actively searched for at LHC
\cite{higgsatlas,higgscms}. Additional ones may also exist  beyond the MSSM, whose particle content  is represented in Table \ref{tab:mssm}.

\section{Some remaining questions, after the 125 GeV boson discovery}

 The standard model constitues a remarkable achievement in the description of the fundamental particle interactions. 
 Even if it is complete, it still 
 leaves many questions unanswered.
 In addition, it  would be presumptuous to imagine 
 that our knowledge of particles and interactions is now complete, without new particles or interactions remaining to be discovered.

\vspace{2mm}
The standard model does not answer many questions, concerning the origin of symmetries and symmetry breaking, the quark and lepton mass spectrum and mixing angles, etc..
Gravitation,  classically described by general relativity, cannot easily be cast into a consistent quantum theory. 
This is why string theories were developed, which seem to require supersymmetry for consistency.
The nature of dark matter 
and dark energy which govern the evolution of the Universe and its accelerated expansion remains unknown, 
as the origin of the predominance of matter over antimatter. 

\vspace{2mm}

Dark matter may be composed, for its main part, non-baryonic, of new particles such as the neutralinos of 
supersymmetric theories, or axions, ...\,.
There may also be new forces or interactions beyond the four known ones, strong, electromagnetic, weak and gravitational.  
And maybe, beyond space and time,  new hidden dimensions, extremely small
(with $L < 10^{-17}$ cm corresponding to $ \hbar c/L > 2$ TeV)  or even stranger, 
 like the anticommuting dimensions of supersymmetry.

\section{More on supersymmetry and superspace}

Supersymmetry enlarges the notions of space and time, already related by the theory of relativity, to a new geometry involving a superspace \cite{revue,sf1,sf2}.
This one possesses new quantum coordinates $\theta_\alpha$  associated with rotations and Lorentz transformations.
In the simplest case the $\theta_\alpha$'s  are the four components of a self-conjugate Majorana spinor $\theta$ called the Grassmann coordinate.
They satisfy anticommutation relations, 
\be
\theta_\alpha\theta_\beta=-\,\theta_\beta\theta_\alpha\,,
\ee
in connection with the spin-$\frac{1}{2}$ character of the Grassmann coordinate $\theta$, 
\vspace{-.4mm}
and  associated fermionic character of spin-$\frac{1}{2}$ particles 
\vspace{-.3mm}
obeying the Pauli exclusion principle.
Each $\theta_\alpha$ has a vanishing square, $\theta_\alpha^{\,2}=0$, very much like two identical spin-$\frac{1}{2}$ fermions cannot be in the same quantum state.

\vspace{2mm}
The notion of point gets replaced by the notion of ``superspace point'',
\be
\hbox{ space-time point} \ \,x^\mu= \left(\!\ba{c} ct\\ \vec x \ea\!\right)\ \ \ \rightarrow \ \ \ (x^\mu,\,\theta_\alpha)\,.
\ee
This means in fact that we shall now consider, in the place of fields $\varphi(x^\mu)$ depending on the time and space coordinates $t$ and $\vec x$, superfields
$
\Phi$ depending on $x^\mu$ and $\theta_\alpha$. It is in practice convenient to rewrite $\theta$ 
as a 2-component complex spinor rather than a 4-component self-conjugate one. A  superfield is then  expressed as
\be
\Phi(x,\theta,\bar\theta)\,.
\ee

\vspace{1mm}

Its expansion in terms of $\theta$ and $\bar \theta$ provides a finite number of component fields, both bosonic and fermionic, in equal numbers.
There are  different types of superfields.  Gauge superfields describe 
massless spin-1 gauge bosons and associated spin-$\frac{1}{2}$ gauginos. 
Chiral superfields describe both spin-$\frac{1}{2}$ and spin-0 fields, corresponding to quarks and leptons with associated squarks and sleptons
for $R_p=-1$ superfields. Or to spin-0 BEH fields with associated higgsinos, for $R_p=+1$ superfields, with the $\theta$ coordinate having $R_p=-1$.

\vspace{2mm}

Supersymmetry transformations act in superspace in a special way reminiscent of a translation for the Grassmann coordinate $\theta$,
combined with a transformation of the space-time coordinate $x^\mu$  involving $\theta$.
Their generators $Q_\alpha$  are in the simplest case the four hermitian 
components of a self-conjugate Majorana spinor $Q$ (or of a 2-component complex spinor).
\,In a schematic way supersymmetry transformations generated by the $Q_\alpha$'s transform, for each superfield, bosonic components into fermionic ones, 
and conversely.
Being of fermionic nature the operators $Q_\alpha$ satisfy anticommutation relations.

\vspace{2mm}

The combination of two infinitesimal rotations around $Ox$ and $Oy$ 
generates an infinitesimal rotation around the third orthogonal axis $Oz$, as expressed by the commutation relation
\be
[J_x,\,J_y]\,=\,i\,\hbar\,J_z\,,
\ee
essential in quantum mechanics. In a somewhat similar way, 
the appropriate combination, now through the anticommutators $\,Q_\alpha Q_\beta +Q_\beta Q_\alpha$, \,of two infinitesimal supersymmetry transformations 
generates a translation in space-time.
This is  expressed by the following (anti)commutation relations in  the supersymmetry algebra
\be
\label{alg}
\left\{ \  
\begin{array}{ccc}
\{ \ Q\, , \, {\bar Q} \ \} \!&=&\! 
- \, 2\,\gamma_{\mu}   P^{\mu} , \vspace {2mm} \cr 
[ \ Q\,, \, P^{\mu} \,] \!&=& \ 0\ \ .
\end{array}  \right.      
\ee
The last commutation relation, $[ \ Q\,, \, P^{\mu} \,]=0$,  expresses that supersymmetry transformations commute with translations.

\vspace{2mm}

Supersymmetry transformations  generated by the spin-$\frac{1}{2}$ operator $Q_\alpha$ change the intrinsic angular momentum of particles, i.e. their spin,
by half a unit (or $\hbar/2$).
 When local supersymmetry transformations are considered this necessitates the possibility of performing local  space-time translations, with parameters $\epsilon^\mu(x)$.
It then requires general relativity, and thus gravitation, to be included in the game, leading to theories of supergravity \cite{sugra,sugra2}, with the spin-2 graviton having for superpartner a spin-$\frac{3}{2}$ gravitino.

\vspace{2mm}
Although the gravitino is a priori coupled with a very small coupling constant $\kappa=\sqrt{8\pi G_N}$, of gravitational strength, it could have a very small mass $m_{3/2}$, depending on the models considered, as in the so-called GMSB models. It could then still play an important role in particle physics  experiments \cite{grav}, possibly appearing as the lightest supersymmetric particle produced at the end of a decay chain, 
carrying away missing energy-momentum.

\section{Relating gauge and BEH bosons, in the supersymmetric standard model}

In supersymmetric extensions of the standard model, the 
quartic interactions between the two doublets $h_1$ and $h_2$  appear as part of the electroweak gauge interactions, their quartic contribution to the potential providing the one of the MSSM
 \cite{R},
 \be
V_{\,\rm quartic}(h_1,h_2)\,=\, \hbox{\large$\dis \frac{g^2+g'^2}{8}$} \ (h_1^\dagger \,h_1-h_2^\dagger\, h_2)^2\,+\, 
\hbox{\large$\dis \frac{g^2}{2}$}\ |h_1^\dagger \,h_2|^2\,.
\ee
The quartic coupling constants are no longer arbitrary as for $\lambda$ in the standard model, but fixed 
in terms of the electroweak gauge couplings
as $(g^2+g'^2)/8$ and $g^2/2$.  This is at the origin of the mass equality (\ref{mhZ}), for conserved supersymmetry.

\vspace{2mm}
This is also at the origin of a mass inequality requiring in the MSSM, at the classical level,  the lightest spin-0 boson $h$ to be lighter than $m_Z\simeq 91$ GeV$/c^2$,
up to radiative corrections which could raise it up to 125 GeV/$c^2$ from large supersymmetry-breaking effects involving heavy stop quarks \cite{martin}.
In the presence of an extra singlet coupled to $h_1$ and $h_2$, however, as in the N/nMSSM, extra contributions to this lightest mass make it easier to reach 125 GeV/$c^2$.

\vspace{2mm}
When the BEH mechanism operates within a supersymmetric theory, it provides {\it massive gauge multiplets} \cite{R}. Each of them 
describes  a massive spin-1 gauge boson, two spin-$\frac{1}{2}$ inos constructed from gaugino and higgsino components, and a spin-0 boson. The latter is actually a BEH boson associated with the spontaneous breaking of the gauge symmetry.
We thus get systematic associations between massive gauge bosons and \hbox{spin-0} BEH bosons, a quite non-trivial feature owing to 
their different gauge symmetry properties and very different couplings to quarks and leptons \cite{epjc,ghgh},
with the general association  
\be
\label{gh2}
Z\   \,\stackrel{SUSY}{\longleftrightarrow }\   \hbox{2 Majorana zinos} \ \,
\stackrel{SUSY}{\longleftrightarrow }\  \hbox{spin-0 BEH boson}\,.
\ee
The spin-0 field partner of the $Z$, expressed in usual notations as
\be
\label{z}
z= \sqrt 2\ \hbox{Re} \  (-\,h_1^0\,\cos\beta+h_2^0\,\sin\beta )\,,
\ee
may also be described in a non-conventional way by the massive $Z$ gauge superfield, expanded as
$\,Z(x,\theta,\bar\theta) = -z/m_Z +\, ... \,-\,\theta\sigma_\mu \bar\theta\    Z^\mu\, + \,...\ $.

\vspace{3mm}

This implies the existence of a spin-0 BEH boson
of mass
\be
m \,\simeq \, 91\  \hbox{GeV}/c^2,\ \ \hbox{\it up to supersymmetry-breaking effects}.
\ee
More precisely the $z$ field  (\ref{z}) corresponds  in the MSSM to a mixing between the lighter and heavier spin-0 eigenstates  $h$ and $H$.
The \hbox{spin-0} partner of the $Z$ may then be identified, in the MSSM or N/nMSSM, ...\,, as the new 125 GeV  boson, up to a mixing angle, possibly small, induced by supersymmetry breaking  \cite{ghgh}.
We also have in a similar way
\be
\label{gh3bis}
W^\pm\ \stackrel{SUSY}{\longleftrightarrow }\ \hbox{2 Dirac winos} \ \stackrel{SUSY}{\longleftrightarrow } \ 
\hbox{
spin-0 boson} \ H^\pm,
\ee
with 
$m_{H^\pm}\!=m_{W^\pm}$ up to supersym\-metry-breaking effects.

\section{Conclusion on scalar bosons and supersymmetry}
 
In addition to superpartners, supersymmetric theories lead to an extended set of spin-0 bosons $H^\pm, H,h,A, ...\,$. \ Some appear as extra states for massive \hbox{spin-1} gauge bosons,
providing a relation between spin-1 mediators of gauge interactions and spin-0 particles associated with symmetry breaking and mass generation.

\vspace{2mm}

Depending on its properties the 125 GeV boson observed at CERN may also be interpreted, up to a mixing angle induced by supersymmetry breaking, as 
{\it the spin-0 partner of the $Z$ under two supersymmetry transformations}, i.e. as a $Z$ that would be deprived of its spin.
 This provides within a theory of electroweak and strong interactions 
 the first example of two known fundamental particles of different spins that may be related by supersymmetry,
\,in spite of their different electroweak properties. 
 
\vspace{2mm}

The next run of LHC experiments may well allow for the direct production of supersymmetric particles.  Even this does not happen, and $R$-odd superpartners were 
 to remain out of reach for some time, possibly due to large momenta along very small compact dimensions,
 supersymmetry could  still be tested in the gauge-and-BEH sector at present and future colliders, in particular through the properties of the new spin-0 boson.

\vspace{3mm}

 \bc
 
 * \hspace{8mm} *
 \vspace{2mm}
 
* \ \ 
 \ec

\end{document}